\documentclass[fleqn,10pt]{wlscirep}
\usepackage[utf8]{inputenc}
\usepackage[T1]{fontenc}
\usepackage{eucal}
\title{A Large-Scale Study of Personal Identifiability of Virtual Reality
Motion Over Time}

\author[1,*]{Mark Roman Miller}
\author[2]{Eugy Han}
\author[2]{Cyan DeVeaux}
\author[3]{Eliot Jones}
\author[3]{Ryan Chen}
\author[2]{Jeremy N. Bailenson}

\affil[1]{Department of Computer Science, Stanford University, Stanford, California, USA}
\affil[2]{Department of Communication, Stanford University, Stanford, California, USA}
\affil[3]{Stanford University, Stanford, California, USA}

\affil[*]{mrmillr@stanford.edu}

\begin{abstract}
In recent years, social virtual reality (VR), sometimes described
as the ``metaverse,'' has become widely available. With its potential
comes risks, including risks to privacy. To understand these risks,
we study the identifiability of participants' motion in VR in a dataset
of 232 VR users with eight weekly sessions of about thirty minutes
each, totaling 764 hours of social interaction. The sample is unique as we are able to study the effect of user, session, and time independently. We find that the number
of sessions recorded greatly increases identifiability, and duration
per session increases identifiability as well, but to a lesser degree.
We also find that greater delay between training and testing sessions
reduces identifiability. Ultimately, understanding the identifiability
of VR activities will help designers, security professionals, and
consumer advocates make VR safer. 
\end{abstract}
\begin{document}

\flushbottom
\maketitle
%
%
\thispagestyle{empty}

\section*{Introduction}

Recently, social virtual reality (VR) has been increasing in popularity.
If it becomes a mainstay in the consumer space, it will be important
to discover, understand, and address the risks associated with its
use. One type of risk is the risk to privacy in these social spaces
through re-identification attacks enabled by the rich nonverbal behavior
traces \cite{Yaremych2019} that VR captures.

Re-identification attacks work by aligning several types of partially
identifying information. For example, knowing only one of a target's
ZIP code, gender, or date of birth is not likely to identify that
target out of the entire United States population. However, these
three data points together do identify about 87\% of United States
residents enumerated in the 1990 census \cite{Sweeney2000}. A similar
pattern holds for web browsers given several browser identifiers including
version, operating system, language, and timezone \cite{Eckersley2010}.
In VR, this re-identification has been demonstrated by
leveraging biometric information or nonverbal behaviors \cite{Pfeuffer2019,Moore2021,Miller2020,Nair2022}.

In judging this threat, it is important to understand how long these
identifying characteristics of individuals last. Some previous work
\cite{Miller2022} indicates a reduction of identifiability over the
long term (e.g., several months), but this work focused on a very
short activity chosen to produce identifying data, and sessions were
not taken regularly to compare identifiability across many different
moments in time. We have collected a large sample size (232 participants)
of a long-duration activity (about 20 minutes per session) over a
long timespan (8 weeks), which has enabled several contributions to
the state of the art:
\begin{itemize}
\item results indicating short samples taken over several sessions are more
identifying than longer samples in fewer sessions 
\item results indicating the delay between training data and testing data
affects identifiability in the range from one to seven weeks 
\item corroboration with previous work \cite{Moore2021} that identifiability
is higher within a session than between separate sessions 
\item a motivation, selection, and justification of a classification model
evaluation metric, \textit{multiclass AUC} \cite{Hand2001} that is
invariant to the number of classes (i.e., individuals) being identified,
producing more effective comparisons across disparate datasets, activities,
and participant pool sizes 
\item a proposal of \textit{body-space coordinates}, a refinement of the
feature space so that the two horizontal dimensions is not based upon
an arbitrary global coordinate system but rather relative to a person's
"forward" direction 
\item demonstration of an inference of gender and ethnicity from motion
data, producing small to medium gains in accuracy over baseline models
\end{itemize}

\section*{Related Work}

First, we give a short description of the risks across VR. Then, we
focus on motion identifiability, first on authentication, a related
task, then on works studying identification proper. We delineate between
authentication and identification primarily by the user's motivation.
Given that the point of authentication is for a system to give a user
elevated privileges, one can design an authentication method assuming
the user is aware of the method and willing to participate. Identification
does not provide access, and so the user is either not aware they
are being identified or perhaps trying to avoid identification.

\subsection*{Risks of Social Virtual Reality}

The growth of social VR has come with a commensurate growth in others
urging caution. Many papers aim to anticipate the risks of VR while
there is time to set norms and design new solutions. Slater and collaborators
\cite{Slater2020} describe several risks including \textit{identity
hacking}, an iteration on catfishing and identity theft where one
user poses as another in a virtual environment.

There are risks to the use of single devices as well. Virtual and
augmented reality devices collect substantial amounts of user data
in order to operate. Modern headsets track motion of several body
parts at about 90 times per second, capturing both deliberate and automatic
behavior \cite{Bailenson2018}. Biometric data itself is available
in several ways. For example, height can be deduced from motion data
\cite{Miller2020} and inter-pupillary distance is available through
the device API \cite{Hosfelt2019}.

Data collection is not limited to the user, though. There are ethical
considerations for bystanders who have their privacy compromised by
virtual or augmented reality systems. In the case of devices tracking
environments, such as always-on cameras in augmented reality systems,
individuals who are not the user often lack data privacy safeguards
and the ability to opt-out of data collection \cite{Roesner2014}.

The operation of virtual and augmented reality requires a massive
amount of data to be processed. It will be important to determine
how to balance the access to this data between confidentiality and
availability.

\subsection*{Motion-based authentication}

One thread of work similar to our work in VR identification is authentication
using motion captured in VR. Motion authentication in VR uses certain
identifiable movements such as throwing a ball \cite{Kupin2019} or
one's natural walking pattern \cite{Shen2018} that can be captured
by the headset and controllers. The ability to utilize these characteristics
as a method for authentication has been demonstrated by several works
\cite{Li2016,Olade2020,Wang2021}. For example, Wang and collaborators
\cite{Wang2021} leverage identifiable head gestures (nodding, turning,
and tilting) in order to develop a biometric authentication method.
Similarly, Li and collaborators \cite{Li2016} utilize head movement
patterns that are formed by a user in response to listening to music
for a similar authentication task.

What makes authentication different from identification is that the
person looking for authentication wants to be authenticated. This
difference allows the authentication designer to ask the person to
be authenticated to remember an action (e.g., how to nod one's head
in response to music) or to perform a specific activity (e.g., throwing
a ball). In the identification problem we pose, this is unrealistic,
as we are investigating identifiability given no interaction with
the target at all.

\subsection*{Identification}

The primary question regarding identification is twofold: what makes
certain situations more amenable to identification than others, and
what about those situations can be remedied? Based upon the work so
far in VR motion identification, there are some threads of work beginning
to appear.

First, the number of users to identify between, which we refer to
as the \textit{classification size}, affects accuracy. Three separate
works explicitly analyze the relationship of classification size to
accuracy \cite{Pfeuffer2019,Miller2020,Wang2021}, and all show effects
that larger problems tend to have lower accuracy, even with the same
training algorithm and underlying dataset. This is simply due to the number of classes increasing the likelihood of data points of another class sufficiently close to data of the class being queried.

Second, the fit of the featurization to the activity also affects
accuracy. Activities that are less unlikely to be organically encountered
while spending time in social VR seem to elicit higher identifiability
accuracies. For example, training using the trajectories taken when
placing blocks in specific places produces 98.6\% accuracy \cite{Olade2020},
standing and watching 360° video produces 95\% accuracy \cite{Miller2020},
throwing a ball at a target produces between 85\% and 91\% accuracy
\cite{Miller2022}, pointing, grabbing, walking, and typing produce
between 48\% and 63\% accuracy \cite{Pfeuffer2019}, and walking and
interacting with virtual objects elicits 37-42\% accuracy \cite{Moore2021}.

While some activities have been known to be identifying, e.g., walking,
what is being discussed here is a fit between the featurization and
the activity. Moore and collaborators intentionally did not innovate
on the featurization in their work in order to make a direct comparison
to previous work \cite{Miller2020}, and the feature set selected
in that work was not good for walking as it was primarily on static
biometric values. On the other hand, that same feature set was unreasonably
effective for 360 videos because participants mostly stood in the
same spot or simply looked around, allowing reliable measurements
of height and hand controller orientation. One may naively think the
most salient and most active measures for the activity (e.g., head
direction in terms of pitch and yaw) would be the most identifying,
but in fact it is the seemingly ancillary data of height and controller
positioning. Though this 'fit' is not clearly defined \textit{a priori},
what is clear is that some tasks are more identifying than others,
some tasks are more commonly performed than others, and those are
not often found together. 

Outside of these three threads on identifiability and its confounds,
there are other works to highlight. There is less work about defense
mechanisms for this data. M. Miller and collaborators \cite{Miller2020}
reduce the training data streams from 18DOF (head and hands position
and rotation) to 3DOF (head rotation only) and reduce accuracy from
95\% to 20\% on a set of 511. Moore and collaborators \cite{Moore2021}
reduce accuracy from 89\% to 32\% on one set of data and 42\% to 13\%
on a second by switching from position-based to velocity-based feature
vectors. Nair, Gonzalo, and Song \cite{Nair2022a} use differential
privacy methods to give plausible deniability to the biometric features
they lay out in \cite{Nair2022}.

Additionally, in the same work \cite{Nair2022} the authors use the
identifying characteristics from the VR data collected by a client-type
attacker to infer demographic information about each user. Compared
to our work, they use a wider array of features, while we focused
specifically on motion. Olade and collaborators \cite{Olade2020}
perform whitebox penetration testing of their authentication method (i.e., describing the inner workings of their authentication method so that it is known to the attacker),
and find that intentional imitation of others, even if the attacker
can watch the target, is not sufficient to fool their authentication
process. Falk and collaborators \cite{Falk2021} demonstrate identifiability
across real-world video and VR-world video by leveraging available
skeleton identification pipelines and innocuous social VR activities. Sabra and collaborators \cite{sabra_exploiting_2023} leverage identified smartphone motion data to link identities to users in VR.
Finally, R. Miller and collaborators \cite{Miller2022} indicate that
moving from one VR headset to another can reduce identifiability.

Finally, some work has been done in demonstrating how much more identifiable
actions can become if a minor degree of influence in the virtual environment
can be permitted. Falk and collaborators demonstrate identifiability
on set a of N=5 when an unprivileged attacker (appearing to the target
as simply another social VR user) asks the target to perform an innocuous
activity like throwing a ball or eliciting an automatic reciprocal
response like waving back \cite{Falk2021}. This influence is even
greater in cases where the attacker can design a gamut of activities
for the target to perform \cite{Nair2022}.

\subsection*{Identification Over Time}

\begin{figure}
\centering \includegraphics[width=5.9in]{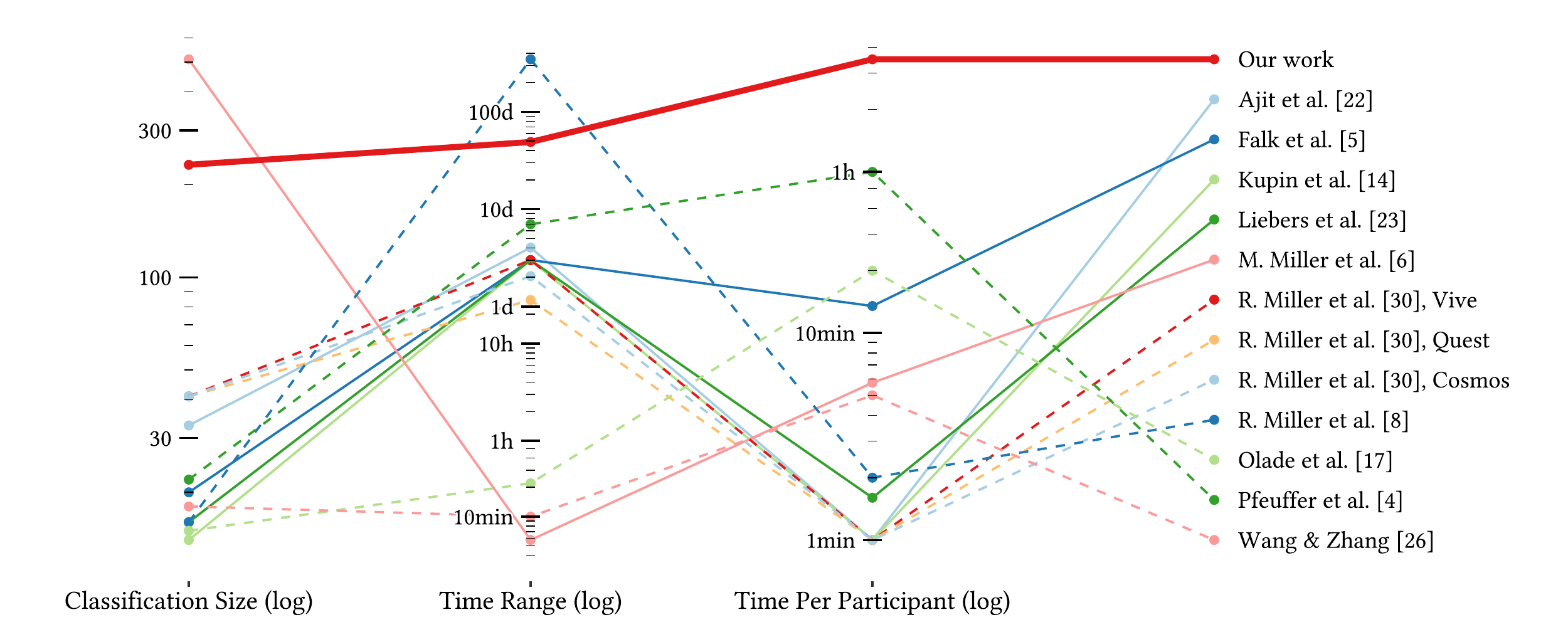} \caption{Parallel coordinates plot of classification size, span of time in
which data was collected, and total duration of data collected per
participant. The current work is the largest or the second-largest on all
dimensions. Note all dimensions are log-scaled in order to better
scale the variation. Papers are given alphabetically by the last name
of the first author.}
\label{fig:comparison} 
\end{figure}

\nocite{Ajit2019}

In this work, we focus specifically on identification over time. The
delay in time between a user's training data (i.e., enrollment) and
a user's testing data (i.e., input) seems to affect accuracy. Works
that have a minimal delay between training and testing have a higher
accuracy than works with longer delays, e.g., a delay of 30 seconds
between sessions and data collected over the span of about an hour
with 98\% accuracy \cite{Wang2021}, no delay and a span of 30 minutes
with 98.6\% accuracy \cite{Olade2020}, no delay and a span of 10-15
minutes with 95\% accuracy \cite{Miller2020}, no delay and a span
of 60 minutes with 89-95\% accuracy \cite{Moore2021}, sessions recorded
on "different days" with 90\% accuracy \cite{Liebers2021}, "at
least three days {[}between{]}" with 63\% accuracy \cite{Pfeuffer2019},
and one week later with 42\% accuracy \cite{Moore2021}. This effect
of time delay is explicitly studied by R. Miller and collaborators
\cite{Miller2022} by combining two sets of data collected up to 18
months apart. They find no effect of delay on short-scale separations
(within 24 hours) or medium-scale separations (comparing delays shorter
than 3 days and longer than 3 days in one analysis, and the same but
for 10 days in a second analysis). On long timescales, which in their
work goes from 7 to 18 months, there were changes in behavior and
a reduction in accuracy, which was not the case in the short- and
medium-term delays. However, the delays were not regularly spaced
and some varied widely in magnitude. It is still an open question
how much and in what situations identifiability changes over time.


In contrast to previous work, we focus on an extremely common social
VR activity, group discussion. This way, an attacker need not elicit
a response nor manipulate the environment. As shown in Figure \ref{fig:comparison},
this work is also on a far larger sample size than most, has more
collected data than any other, and collected that data for a longer
duration than most. Additionally, there is regular spacing of data
collection periods. This dataset is uniquely posed to answer the next
round of questions about identifiability.

\section*{Methods}

This work reports on data collected as part of the Stanford Longitudinal
VR Classroom Dataset over the course of two periods of data collection
of classroom immersive VR \cite{Han2023}. Students met in small groups ranging from
two to 12, and consented to have their verbal, nonverbal, and performance
continually tracked during each course, typically eight weekly sessions
which lasted about 30 minutes per session. In addition, each student
provided self-report data about their experience after each session (see \cite{Han2023} for a detailed description). The current paper utilizes
previously unreported data from the dataset, and focuses on identifiability
of this nonverbal motion data. These studies from which we report
data were each run using a social VR platform called ENGAGE. Each
data collection period consisted of its own participant pool and conditions.
Participant consent went through a rigorous process, approved by
two separate organizations within the university. Moreover, there
was a 3rd party arbiter who oversaw data collection during the course,
and students had an interactive, hour-long discussion of the study
procedures and data collection before deciding to consent. Data recorded
included position and rotation of each participant's headset and hand
controllers.

\subsection*{Threat model}

It is important to establish the kind of threat under study. In previous
work on VR and identifiability, there are two dimensions upon which
researchers have categorized threats. First, there is the question
of what data is available to the attacker. Nair and collaborators
\cite{Nair2022} delineate between hardware-level attackers that have
access to firmware, client-level attackers that have access to the
headset APIs, server-level attackers that have access to the telemetry
data sent to the servers and 'unprivileged user' attacker which is
another VR system partaking in the same social virtual world. Along
this dimension, we focus on the unprivileged user.

The second aspect of space of threat models is the capability of the
attacker to influence the behavior of the participant, and the extent
to which this can be done. For example, is the attacker designing
a virtual world \cite{Nair2022}, are they another user that is interacting
with the user \cite{Falk2021}, or do they wish not to interact with
the target entirely? In our work, we focus on no interaction at all.
This may occur because the attacker is working with previously-collected
data, does not want to be vulnerable in the virtual world, or has data
collected at scale and cannot interact with each user.

This threat model is selected because it is the least privileged attacker.
Therefore, findings based on this work are likely to be applicable
to all attacks leveraging motion data. It also sets a baseline on
threat for all these other conditions. Finally, there are some cases
in which this may be the mode of an attacker, e.g., large-scale surveillance
where individuals are not queried directly, re-identification attacks
where actions are stored for a period of time before being queried,
or any other situations in which the attacker does not which to have
any direct interaction with the target. Note that this threat model
is quite different from the traditional authentication threat model
in which a user attempts to gain unauthorized access by posing as
another user.

We have also selected the activity, group social discussion, to be
representative of activities one may encounter in social VR. This
is in contrast to activities like throwing a ball that may not be
encountered regularly. This activity also makes the findings more
ecologically valid.

\subsection*{Apparatus}

Participants were provided with Meta Oculus Quest 2 headsets (503g)
and two hand controllers (126g) for use at home or another personal
environment. The Quest 2 headsets are standalone head mounted displays
with 1832 x 1920 resolution per eye, 104.00° horizontal FOV, 98.00°
FOV, 90Hz refresh rate, and six degree-of-freedom inside-out head
and hand tracking. In the first round of data collection, two participants
opted to participate with owned personal headsets (both PC-based Valve
Index).

\begin{figure}
\centering \includegraphics[width=4in]{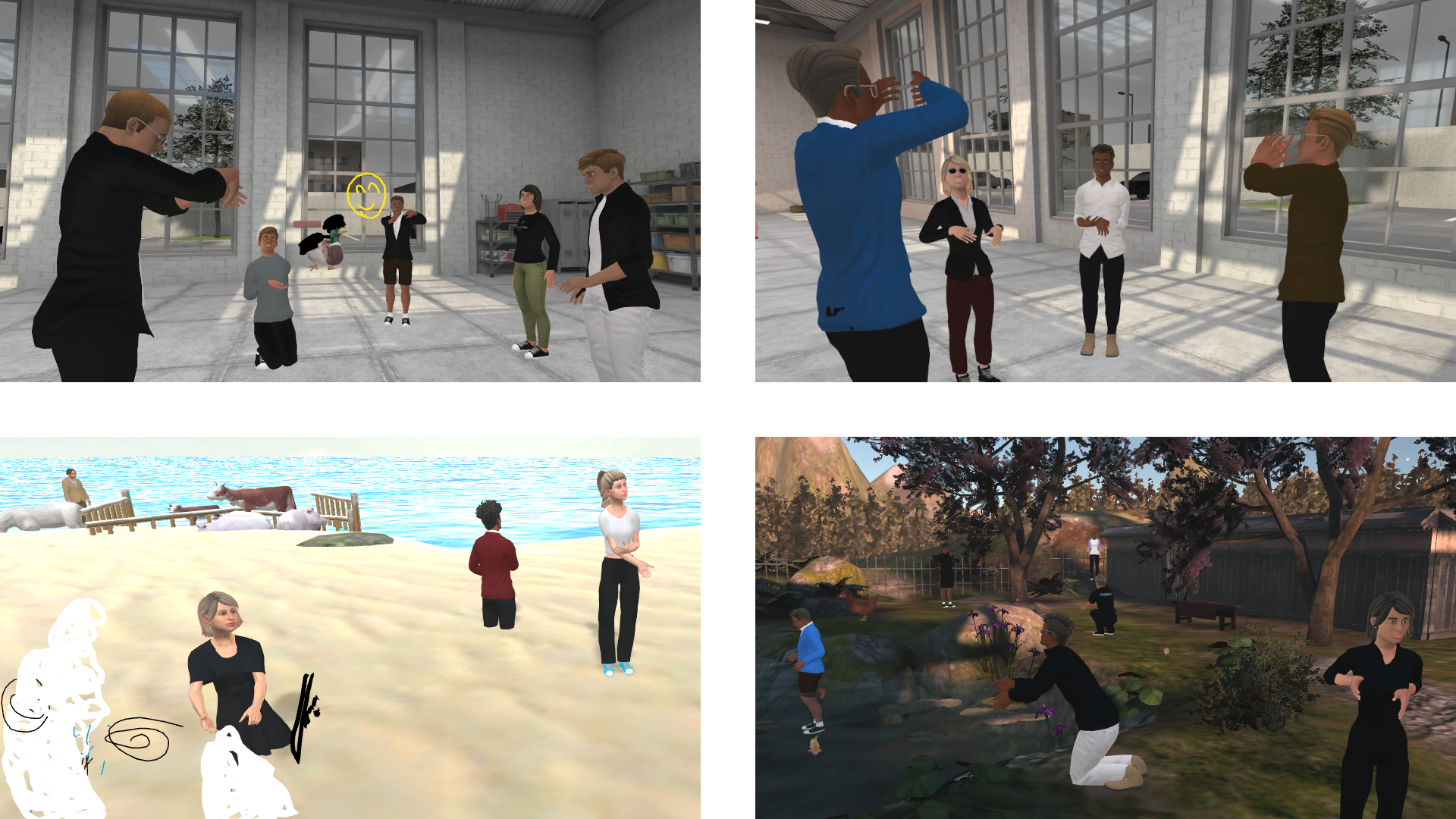} \caption{Participants performing discussion activities in the VR environment.
In the top left panel, participants illustrate the environmental impacts
of an oil spill with a duck model covered in black smudges representing
oil. In the top right panel, several students discuss the experience
of the headset using hand gestures in front of their faces.
In the bottom left panel, the participants used 3D drawing to illustrate the water cycle as well as 3D models to show livestock in fenced-in environments. In the bottom right panel, participants are spreading out into different corners of the space and brainstorming on how to visualize their ideas to the assigned prompt.}
\label{fig:engage_screenshot} 
\end{figure}

The software in use was the ENGAGE virtual communications platform,
versions 1.7 through 2.0.1, produced by ENGAGE PLC. The virtual environments
in which the participants met were private (password restricted).
In the first round of collection, all participants met in the same
``Engineering Workshop'' room. In the second round of data collection,
participants met in one of 192 uniquely-built environments each week.
These environments differed in size of moving area and height. Figure
\ref{fig:engage_screenshot} shows screen captures of anonymized virtual
students in action.

\subsection*{Participants}

There were a total of 232 participants in the study across the two
periods of data collection ($n_{1}=86$), ($n_{2}=146$). Participants
were university students enrolled in one of two 10-week courses about
VR. While all students who were part of the course took part in all
the VR activities, only those who consented to participate in the
study had their data included in the study. Of the 101 students in
Period 1 and 171 in Period 2, 93 and 158 consented to participate
in the study, respectively. Additionally, 6 students in Period 1 and
12 students in Period 2 dropped the class after consent was collected
and before data collection began.

In Period 1 (Female = 30, Male = 47, Other = 2, declined or did not
answer = 7), participants were between 18 and 58 years old (M = 22.3,
SD = 5.2; $n_{18\text{--}23}$ = 68, $n_{24\text{--}29}$ = 7, $n_{30\text{--}34}$
= 3, $n_{35\text{--}39}$ = 1, $n_{55\text{--}59}$ = 1, $n_{declined}$
= 6 and identified as African American or Black ($n$ = 11), Asian
or Asian American ($n$ = 30), Hispanic or Latinx ($n$ = 9), Middle
Eastern ($n$ = 1), White ($n$ = 21), more than one race ($n$ =
5), or declined to or did not respond ($n$ = 9). Participants had
varying levels of experience with VR, with 41 (51.2\%) having never
used VR before. Prior to the course, 38 participants were not familiar
with anyone in their discussion group, and others reported knowing
one ($n_{1}$ = 13) or more members ($n_{2}$ = 12, $n_{3}$ = 1,
$n_{4}$ = 2, $n_{5}$ = 2).

In Period 2 (Female = 59, Male = 79, declined or did not respond =
4), participants were between 18 and 49 years old (M = 20.9, SD =
2.8; $n_{18\text{--}23}$ = 133, $n_{24\text{--}29}$ = 4, $n_{45\text{--}49}$
= 1, $n_{declined}$ = 4 and identified as African American or Black
($n$ = 12), Asian or Asian American ($n$ = 47), Hispanic or Latinx
($n$ = 8), Indigenous/Native American, Alaska Native, First Nations
($n$ = 2), Middle Eastern ($n$ = 1), Native Hawaiian or other Pacific
Islander ($n$ = 5), White ($n$ = 41), more than one race ($n$ =
19), a racial group not listed ($n$ = 1), or declined to or did not
respond ($n$ = 2). Participants had varying levels of experience
with VR, with 50 (36.2\%) having never used VR before. Prior to the
course, 67 participants were not familiar with anyone in their discussion
group, and others reported knowing one ($n_{1}$ = 36) or more members
($n_{2}$ = 10, $n_{3}$ = 4, $n_{4}$ = 5, $n_{5}$ = 1, $n_{7}$
= 1).

\subsection*{Procedure}

Students opted in to the experiment at the beginning of the course
with a consent form approved by the Stanford University institutional review
board (IRB) under protocol IRB-61257, and the Stanford University Student Oversight Committee. This
IRB process required that researchers and course staff did not know
which students opted in as participants in the experiment until after
the course finished, and an external third-party arbitrator controlled the consent process, so that there would be no plausible appearance
of coercion to participate in the study. This also implied that all
students were recorded in this study, and data was filtered out from
non-consenting participants after the course finished. Furthermore,
participants were not compensated because they performed the same
activities regardless of whether their data was used. Upon the start
of each session's recording, the system gave a visual notification that
recording was taking place. Before consent was given, one of the authors gave a 30-minute lecture on data tracking privacy, and the pros and cons of consenting in the study, and gave the students the opportunity to ask questions about the study. All experimental protocols were approved by the Stanford University IRB, all participants' informed consent was obtained, and all methods were carried out in accordance with relevant guidelines and regulations.

Weekly activities varied, but included large-group discussion on current
readings, discussion in pairs or triads on course material, and VR
building activities. Sessions were led by a researcher, who was part
of the teaching team of the course. In all virtual environments, participants
were able to walk/teleport freely, create 3D drawings, write on personal
whiteboards/stickies, add immersive effects/3D objects, and display
media content. There was a library of about one thousand virtual objects
available for participants to create, move, organize, and delete in
the virtual spaces. The platform accommodated use of 3D audio, which
allowed for splitting off into smaller groups without audio overlap.
Sessions took place eight times over the course of eight to nine weeks,
and the duration was about thirty minutes per session.

\subsection*{Data}

The highest level of data organization was the \textit{dataset}, which
was either dataset 1, collected in summer 2021, or dataset 2, collected
in fall 2021. The next levels of organization are the week and the
section. The \textit{week} indicates which week of eight the data
is collected from. The \textit{section} was the group and time participants
met for discussion, which is orthogonal to week. In dataset 1, there
were eight sections, and in dataset 2, there were twenty-four. Each
participant took part in only one section per week. On average, there were 6.57 participants per section, with numbers of participants ranging from two to twelve. Usually, a participant
attended the same section week to week, but there were some exceptions.
A \textit{session} is one participant's data for one week. Each session
lies entirely within one and only one section. In total, there were
1,683 sessions with an average length of 27.2 minutes (SD = 11.2 min).

During a session, the data was collected at 30Hz and consisted of
four tracked points. Three were the traditional head, left hand, and
right hand, and the fourth is the 'root', the relationship between
the participant's physical space and the virtual space. This was used
in cases when a participant translated or rotated their position with
a UI control (e.g., teleporting by pointing and clicking, tapping
the controller joystick left to rotate left by 15 degrees). As we
were interested in an attack made by a malicious user and not a compromised system, we computed
the avatar's head and hands positions as visible to another user,
which is relative to the virtual spaces' coordinate system.

The coordinate system of the data itself follows the convention system
used by the Unity game engine; namely, a left-handed coordinate system
with Y upwards, Z forwards, and X rightwards, and that with intrinsic
rotations in the order of yaw (Y), pitch (X), and roll, (Z), where
positive values indicate a left-handed rotation relative to the positive
direction along the axis.

\subsection*{Feature Engineering}

The feature set we used consisted of 840 \textit{streams} that were subset and
summarized in various ways. Some of these streams were defined in
terms of body-space coordinates, which is described below.

\subsubsection*{Body-space Coordinates}

\label{bsc}

The registration of a coordinate system is often not amenable to moving,
flexible, and diverse human bodies. Over time, different coordinate
systems have been developed for specific purposes, such as the anatomical
planes (coronal, sagittal, transverse) for medical terminology. For
the purposes of our work specifically and of VR more generally, we
propose a coordinate system that synthesizes the global vertical axis
with horizontal axes relative to the headset's forward direction.

\begin{figure}
\centering \includegraphics[width=5.9in]{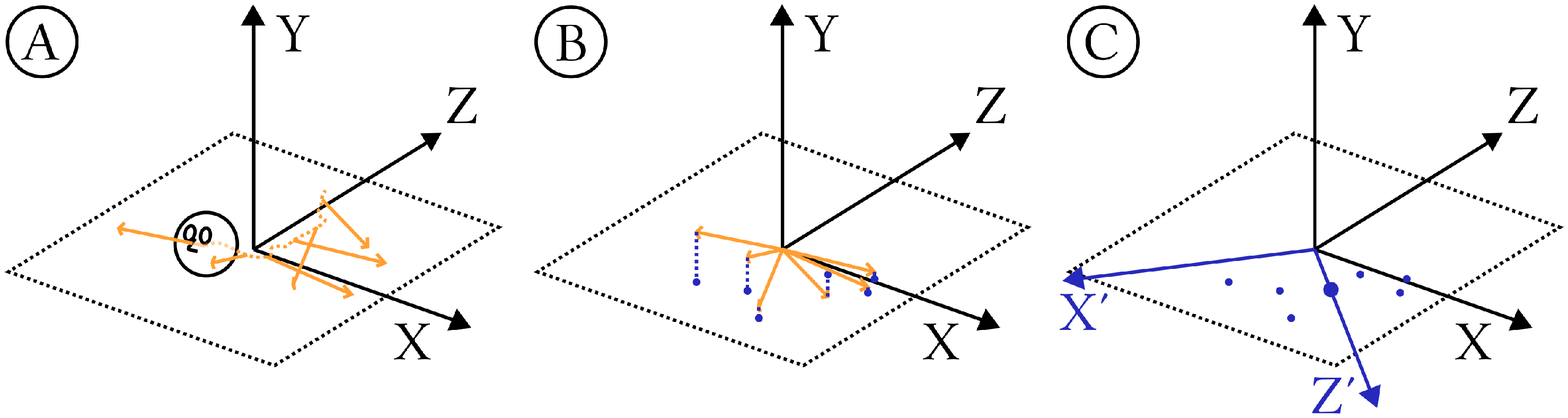}
\caption{Body-space coordinates. In Panel A, a span of one person's motion
is shown in space. The orange arrow represents the direction the headset
is pointing. In Panel B, all these direction vectors are translated
with initial points at the origin. The blue dots are the projection
of the orange forward vectors onto the XZ (horizontal) plane. Panel
C denotes these blue dots, the large blue dot showing the mean of
these points, and the Z' and X' axis denoting forward and rightward,
respectively.}
\label{fig:bsc} 
\end{figure}

In this section and next, we designate $\mathbf{p}_{\alpha}[f]$ a
$3\times1$ vector specifying the position of object $\alpha$ at
frame $f$ in global coordinates, where $\alpha=h$ for head, $l$
for left controller, and $r$ for right controller. Each dimension
can individually be accessed as $px_{\alpha}[f]$, $py_{\alpha}[f]$,
and $pz_{\alpha}[f]$. We also define $\mathbf{R}_{\alpha}[f]$ as
a $3\times3$ matrix specifying the rotation of object $\alpha$ at
frame $f$. Additionally, we define $yaw(\mathbf{R})$, $pitch(\mathbf{R})$,
and $roll(\mathbf{R})$ as functions receiving a rotation $\mathbf{R}$
and returning a single real number indicating the corresponding Tait-Bryan
angles using the Unity game engine rotation conventions. All these
expressions implicitly refer to one session.

The construction of the body-space coordinate system is illustrated
in Figure \ref{fig:bsc} and described as follows: while the vertical
(Y) direction remains vertical in this coordinate system, one of the
horizontal directions (in our convention, Z) is defined to be forward
relative to the participant's body. We operationalize the forward
direction as the horizontal direction of the mean of the head's forward
vector over the course of a span $\Delta f$ of frames $F$ centered
on a query frame $f$. In our work, the span is $\Delta f=\pm3s$.
The projection onto the horizontal plane downweights directions facing
upward or downward, which we judge is appropriate. The rightward direction
is then defined as the cross product of upwards with forwards so that
it is orthogonal to both. Mathematically speaking, if $F$ is the
set of frames over which to calculate a forward vector, then the transformation
$\mathbf{R}_{bsc}$ to the body-space coordinate system is calculated
as follows:

\begin{align*}
F[f] & =\{f':f-\Delta f\leq f'\leq f+\Delta f\}\\
\mathbf{v}[f] & =\frac{1}{|F[f]|}\sum_{f'\in F}\begin{bmatrix}1 & 0 & 0\\
0 & 0 & 0\\
0 & 0 & 1
\end{bmatrix}\mathbf{R}_{h}[f']\begin{bmatrix}0\\
0\\
1
\end{bmatrix}\\
\theta[f] & =atan2(v_{x}[f],v_{z}[f])\\
\mathbf{R}_{bsc}[f] & =\begin{bmatrix}\cos(\theta[f]) & 0 & -\sin(\theta[f])\\
0 & 1 & 0\\
\sin(\theta[f]) & 0 & \cos(\theta[f])
\end{bmatrix}
\end{align*}

The intuition behind the value of body-space coordinates straightforward:
often the identifiable features of one's motion are invariant to rotations
within the horizontal plane. For example, when one crosses their arms
while facing northwest, then ten minutes later does the same pose
but facing south, the relative position to each other will be similar
relative to the body-space coordinate system, but not relative to
the global coordinate system. In regards to the question at hand,
this would mean the body-space coordinate system is likely to be more
effective at separating one's pose from another's than the global
coordinate system would be. To our knowledge, this has not been done
in the study of identifiability of VR data, because in previous work
there is often a clear 'forward' direction, and that forward direction
can be either enforced or maintained in the course of an experiment
\cite{Miller2020,Miller2022,Olade2020}. The value of body-space coordinates
is further discussion in the subsection on identification features.

\subsubsection*{Features}

In all, there were 42 streams (positions and speeds) each summarized
in 20 ways for each sample, leading to a total of 840 features.

The first nine of the 42 streams were kept from our starting model,
which was originally used by \cite{Miller2020} and re-used in \cite{Moore2021}.
These were vertical position, roll, and pitch for each of the head
and the left and right hand controllers. The subsection on identification features
details the value of dropping the other three data streams, specifically,
yaw, X position, and Z position. In the notation given above, the
first nine streams were

\[
py_{h}[f],\ py_{l}[f],\ py_{r}[f],\ pitch(\mathbf{R}_{h}[f]),\ pitch(\mathbf{R}_{l}[f]),\ pitch(\mathbf{R}_{r}[f]),\ roll(\mathbf{R}_{h}[f]),\ roll(\mathbf{R}_{l}[f]),\ roll(\mathbf{R}_{r}[f])
\]

The next nine streams are defined in terms of body-space coordinates.
First, each point is calculated as the difference between tracked
objects $\alpha$ and $\beta$ by $\mathbf{d}_{\alpha\beta}[f]=\mathbf{p}_{\alpha}[f]-\mathbf{p}_{\beta}[f]$
in a method similar to \cite{Miller2022a}. Then, this difference
is transformed to body-space coordinates $\mathbf{b}_{\alpha\beta}[f]=\mathbf{R}_{bsc}[f]\mathbf{d}_{\alpha\beta}[f]$.
In this body-space coordinate system, y refers to difference upward,
x to difference rightward, and z to difference forward. All three
dimensions are used in all three pairs, producing nine streams:

\[
bx_{h}l[f],\ bx_{h}r[f],\ bx_{r}l[f],\ by_{h}l[f],\ by_{h}r[f],\ by_{r}l[f],\ bz_{h}l[f],\ bz_{h}r[f],\ bz_{r}l[f]
\]

The third set is also nine streams, but they are defined on speed
rather than position as before. Specifically, one-frame changes in
position are denoted $\mathbf{v}_{\alpha}[f]=\mathbf{p}_{\alpha}[f]-\mathbf{p}_{\alpha}[f-1]$.
Three types of speeds are derived, one that indicates total motion
$|\mathbf{v}_{\alpha}[f]|$, one that indicates horizontal motion
$|\mathbf{v}_{\alpha}[f]|_{H}=\sqrt{vx_{\alpha}[f]^{2}+vz_{\alpha}[f]^{2}}$,
and one that indicates vertical motion $|\mathbf{v}_{\alpha}[f]|_{V}=|vy_{\alpha}[f]|$.
This results in nine streams, specifically:

\[
|\mathbf{v}_{h}[f]|,\ |\mathbf{v}_{l}[f]|,\ |\mathbf{v}_{r}[f]|,\ |\mathbf{v}_{h}[f]|_{H},\ |\mathbf{v}_{l}[f]|_{H},\ |\mathbf{v}_{r}[f]_{H},\ |\mathbf{v}_{h}[f]|_{V},\ |\mathbf{v}_{l}[f]|_{V},\ |\mathbf{v}_{r}[f]|_{V}
\]

Finally, on the speed of the body-space difference vectors $\mathbf{v}_{\alpha\beta}[f]=\mathbf{b}_{\alpha\beta}[f]-\mathbf{b}_{\alpha\beta}[f-1]$,
one can compute the same total, horizontal, and vertical motions,
but it is meaningful to also compute the difference along both the
forward $|\mathbf{v}_{\alpha\beta}[f]|_{F}=|vz_{\alpha\beta}[f]|$
and rightward $|\mathbf{v}_{\alpha\beta}[f]|_{R}=|vx_{\alpha\beta}[f]|$
directions. Therefore, there are a total of 15 streams of this type:

\begin{gather*}
|\mathbf{v}_{hl}[f]|,\ |\mathbf{v}_{hr}[f]|,\ |\mathbf{v}_{lr}[f]|,\ |\mathbf{v}_{hl}[f]|_{H},\ |\mathbf{v}_{hr}[f]|_{H},\ |\mathbf{v}_{lr}[f]|_{H},\ |\mathbf{v}_{hl}[f]|_{V},\ |\mathbf{v}_{hr}[f]|_{V},\ |\mathbf{v}_{lr}[f]|_{V},\\
|\mathbf{v}_{hl}[f]|_{F},\ |\mathbf{v}_{hr}[f]|_{F},\ |\mathbf{v}_{lr}[f]|_{F},\ |\mathbf{v}_{hl}[f]|_{R},\ |\mathbf{v}_{hr}[f]|_{R},\ |\mathbf{v}_{lr}[f]|_{R}
\end{gather*}

The featurization for any given session is a collection of 840-entry
vectors computed for a set of frames regularly spaced at one second
intervals. The 840 features are computed by selecting each one of
the 42 streams and computing each of five different summary statistics
(mean, median, maximum, minimum, standard deviation) of each of four
varying window sizes (1s, 3s, 10s, 30s) centered on the selected frame.
On the edges, where a full window frame was not available, the frame
was not used.

\subsection*{Model}

The model we have selected is random forest implemented in R \cite{RLang2021}
with the package \texttt{ranger} \cite{rangerPkg2017}. The choice
of random forest was to balance simplicity with expressiveness and
was used in previous work \cite{Miller2020,Moore2021}. Most settings
for the creation of the random forest were the defaults, in particular,
no limit to node depth, a minimum node size of one, number of variables
to try as the square root of total variables. The one custom parameter
was training only 30 trees on a sample of 100,000 entries from the
entire database, with this resample-and-grow process repeated 20 times.

The predictions were made per session by taking the entire session
of motion data, computing features, and then aggregating the votes
across all 600 trees across all samples. We interpreted this distribution
as a probability estimation for the classification of the session
as a whole, in line with previous work \cite{Miller2020} and consistent
with other uses of random forests \cite{niculescu-mizil_predicting_2005}.

\subsection*{Evaluation}

We provide several evaluation metrics for these models. Our primary
metric is multiclass AUC \cite{Hand2001} which addresses the dependency
of accuracy as the number of classes in the classification problem
varies. For the sake of interpretability, we include accuracy, and
for comparison and synthesis with previous work, we use accuracy limited
to a N-class testing set.

\subsubsection*{Multiclass AUC}

To our knowledge, no work in the space of user identification with
VR data has used multiclass AUC. Because it addresses the effect of
classification size on accuracy, we give a short description and justification
of its use in enabling future comparisons across studies with varying
numbers of classes.

Identification-focused works \cite{Pfeuffer2019,Miller2020,Moore2021,Miller2022}
almost exclusively use accuracy for the model's evaluation metric.
The benefits of accuracy as a metric include its ease of interpretation
and its directness to the question at hand - a less accuracy model
is obviously less identifiable, and vice versa. However, accuracy
does vary significantly as the number of classes varies, even for
the same data distributions and identification processes, as evidenced
by multiple works \cite{Pfeuffer2019,Miller2020,Wang2021}. Intuitively,
this is true - it is easier to guess who is walking up the stairs
in an apartment with two other people than a house of ten. This effect
of the number of classes on accuracy can make synthesis of findings
across works difficult, as the classification can vary as much as
two orders of magnitude (e.g., 5 in \cite{Wang2021} to 511 in \cite{Miller2020}).

Our criteria for an evaluation metric that addresses this issue is
that it produces the same value regardless if it is computed upon
the full set of classes, or computed as the average of randomly chosen
subsets of classes of any size. A formal mathematical description
of this problem is given in the supplementary material. To solve this
problem and enable comparisons across analyses with varying numbers
of classes, we choose our primary evaluation metric to be \textit{multiclass
AUC}, defined by Hand and Till \cite{Hand2001}. Multiclass AUC can
be described as the average of the pairwise separability between classes. 

Hand and Till note that this metric weights the separability of each
pair of classes equally regardless of the number of samples in the
classes, which may not be appropriate if priors are to be taken into
account. Additionally, this is not an estimate of the accuracy attained
by the same training process upon a smaller data set constructed in
the same class-reduction process, but is instead an estimate based
upon the model after training.

\subsubsection*{Accuracy limited to an $N$-class testing set}

While multiclass AUC is a good multiclass evaluation metric for future
work, there are no works in this space that currently use it. In order
to allow comparisons to be drawn from this work to previous work,
we define accuracy limited to N-classes. This metric may be narrated
as a prediction task in which there is a model and a set of N potential
classifications, a subset of all the classifications the model could
make. First, the model proposes its classification, and if the classification
is outside this subset, the model is asked to provide its next best
classification. This process only ends when the model gives a predicted
classification within the set of potential classifications. A formal
description of this process and metric is given in the supplemental
material. This accuracy is then comparable to accuracy of a similar
size identification set.

\section*{Results}

There are several analyses performed. First, we estimate the identifiability
of participants in this dataset using both within- and between-session
methods. From there, we study the effect of time on identification,
first through the duration of the training data, and second through
the delay between training and testing.

\subsection*{Identification}

\label{identification}

The simplest and most prominent question is identification. As previous
work has indicated that the delay between training and testing data
sets can influence identification \cite{Miller2022,Moore2021}, we
report two separate analyses. The first analysis uses a train/test
split at the scale of weeks such that the first six weeks of data
are the training set and the final two weeks are the testing set.
This is called the \textit{between} sessions split. The second analysis
splits training and testing segments within each session, such that
80\% of each session is used for training data, and 20\% is used for
testing data, except for one minute as a buffer between segments.
This is called the \textit{within} session split. 
Results are given in Table \ref{tab:ident} that show each accuracy
metric for the two splits across the dataset 1, dataset 2, or the
combined dataset.


\begin{table}
\caption{Evaluation of Identification Models by Train-Test Split and Dataset}
\label{tab:ident} %
\begin{tabular}{lllll}
\toprule 
Split  & Dataset  & Accuracy  & Multiclass AUC  & 30-Class Accuracy\tabularnewline
\midrule 
 & Combined (C=232)  & 31.68\%  & 85.48\%  & 51.71\% \tabularnewline
Between  & Dataset 1 (C=86)  & 45.19\%  & 86.86\%  & 55.34\% \tabularnewline
 & Dataset 2 (C=146)  & 32.39\%  & 86.37\%  & 50.02\% \tabularnewline
\midrule 
 & Combined (C=232)  & 67.15\%  & 98.43\%  & 85.92\% \tabularnewline
Within  & Dataset 1 (C=86)  & 82.30\%  & 98.60\%  & 89.29\% \tabularnewline
 & Dataset 2 (C=146)  & 69.36\%  & 98.28\%  & 84.46\% \tabularnewline
\bottomrule
\end{tabular}
\end{table}

The most dramatic difference for all three evaluation metrics is whether
the split is within-session or between-session. This corroborates
previous work \cite{Moore2021,Miller2022} that the delay between
training and testing influences identifiability. Second, in both cases
the accuracy for dataset 1 is substantially larger. While multiclass
AUC also has slight increases in conditions with dataset 1, it is
much less pronounced, both in an absolute scale and in logit units.
This result highlights the importance of accounting for classification
size. However, it should be noted that identification size is not
the only determinant of accuracy. There is a much smaller difference
in accuracy between the dataset 2 and the combined data (+60\% size
difference) than there is between dataset 1 and dataset 2 (+70\% size
difference).

\subsection*{Identification over time}

Motivated by R. Miller et al. \cite{Miller2022} and the finding in
the subsection above, we investigated the effect
of duration and delay on identification. To study duration, we vary
the number of separate sessions in the training set and the training
time per session to investigate its effect on multiclass AUC for both
within and between session splits. To study delay, we vary the weeks
upon which a model is trained and tested while keeping the duration
the same.

\subsubsection*{Duration\label{subsec:Duration}}

In the first analysis, the train-test split was performed by first
randomly selecting a set of training sessions of size at most 1, 2,
4, or 7 for each participant, presuming at least one session remains
for testing. For example, in the case seven sessions were requested
but a participant only took part in six, five of those six were used
for training and one was held out for between-sessions testing. Of
the selected training sessions, spans of time for training and within-sessions
testing were chosen. First, a five-minute span was marked out for
testing, and one-minute spans adjacent to the testing span were marked
off as buffers. Then, if there was enough space remaining on either
side of the testing span, the training span was one continuous block.
Otherwise, the training span was two separate blocks that totaled
the requesting time. In the case there was not enough time in a session,
the entire length of time other than testing and buffer was used for
training. For example, if there was a 32-minute recording with 30
minutes requested, there would be 25 minutes for training, five minutes
for testing, and one minute on each side of the testing data as a
buffer. This means the average training span for each of the 1, 3,
10, and 30 minute conditions were durations of 1:00, 2:59, 9:39, and
22:32 respectively. Sessions shorter than eight minutes total were
dropped from this analysis.

\begin{figure}
\centering \includegraphics[width=5in]{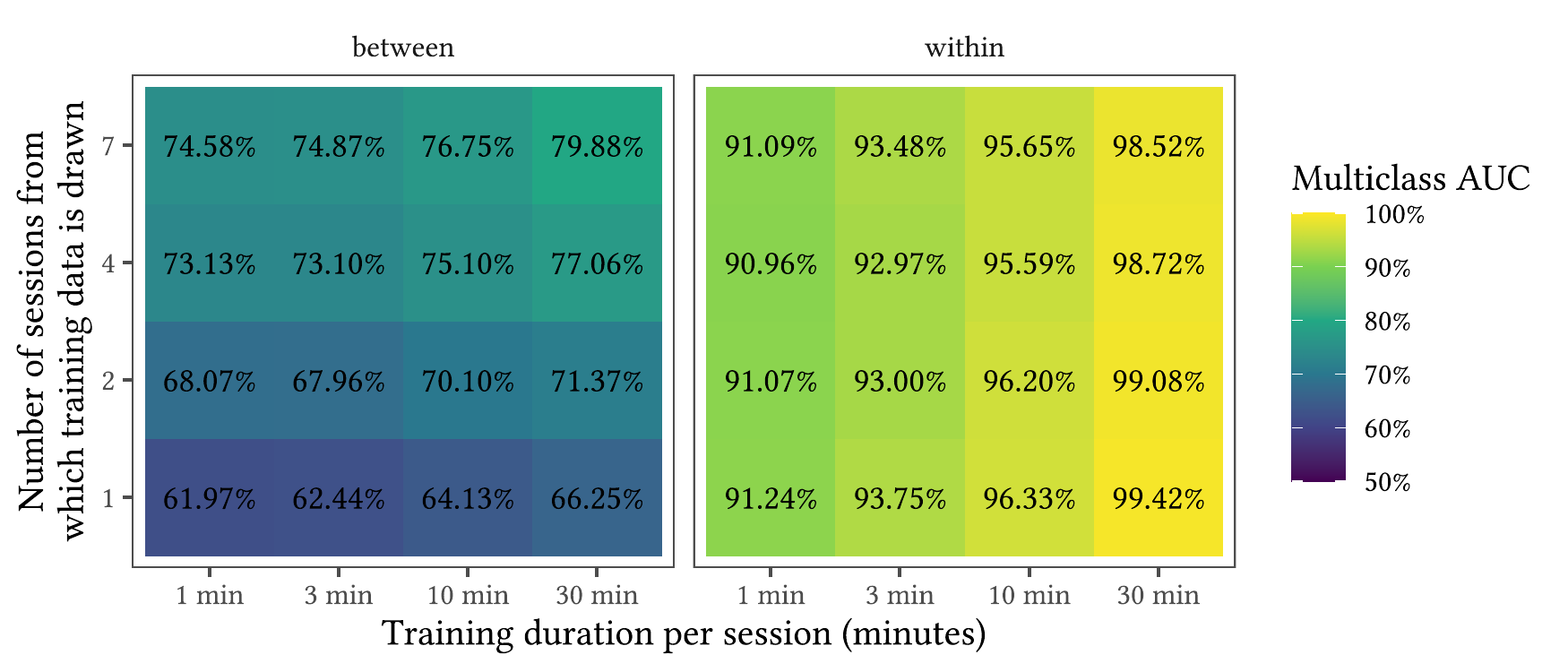} \caption{Number of sessions and duration of each session affect accuracy. Two
panels shown side-to-side indicate whether the comparison is drawn
between sessions or within the same session. The number of sessions
is the y-axis, and the training duration per session is the x-axis.
The panels are colored indicating accuracy, with yellow as a higher
accuracy.}
\label{fig:timing_duration} 
\end{figure}

The reported result is the average multiclass AUC across 10 Monte
Carlo resamplings. For the sake of training time, we reduced the number
of trees to 5 and the number of sub-samples to 3. With this change
in the training process, the reduction in the training set, and the
reduction in the test set to five minutes, we expect to account for
the differences in multiclass AUC between the 30 minute, 7 session
(top right corner, between panel) training and 5 minute testing (i.e,
the results in Figure \ref{fig:timing_duration}) compared to the
full duration, 6 session training set with a full session testing
set (i.e., the results in table \ref{tab:ident}). Results are given
in Figure \ref{fig:timing_duration}.

Again, the most dramatic difference in multiclass AUC is in the delay
(between vs. within) distinction, even so far that training on simply
one minute from the same session as the testing set provides a better
multiclass AUC than thirty minutes on all other sessions available
to the model. Looking specifically at the between-sessions case, both
more sessions and more training data per session produces higher multiclass
AUC scores. Of the two, the number of sessions more dramatically affects
the multiclass AUC, considering the ranges chosen in this study. In
particular, training on one minute from four sessions produces a higher
multiclass AUC than training on thirty minutes from only two sessions
($t(15.06)=3.194$, $p=0.006$, 95\% CI = {[}0.5\%, 2.9\%{]}). In
the within-session data, more time is beneficial to multiclass AUC,
but more sessions very mildly reduces multiclass AUC. This may be
because the variation between sessions does not aid the model but
rather misleads it. 

\subsubsection*{Delay}

\begin{figure}
\centering \includegraphics[width=5in]{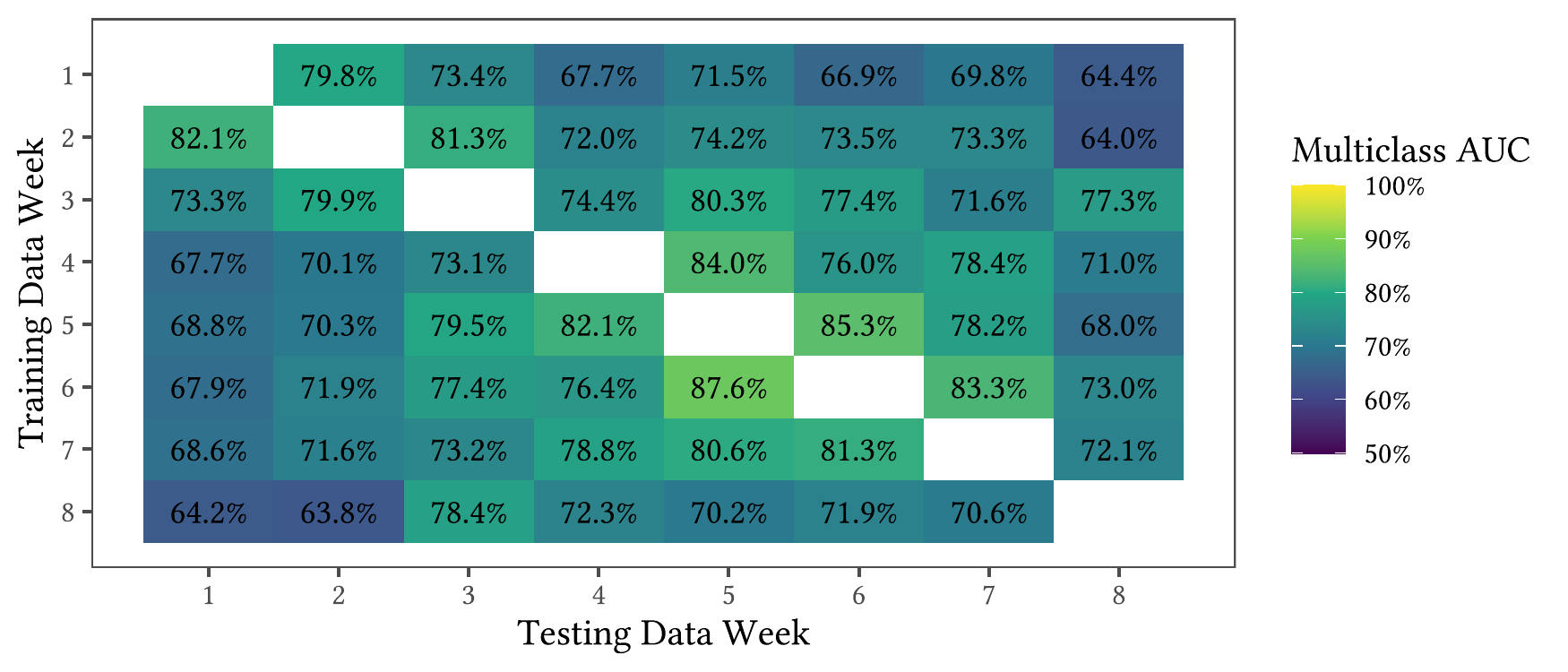} \caption{Separating the training and testing sets by larger time reduces accuracy.
The x-axis and y-axis are the testing and training weeks, respectively.
The panels are colored indicating accuracy, with yellow as a higher
accuracy. Note a trend that higher accuracy is along the diagonal
(i.e., minimal delay).}
\label{fig:timing_delay} 
\end{figure}

The second analysis keeps the training and testing durations the same
but varies the delay between these moments. The multiclass AUCs reported
in Figure \ref{fig:timing_delay} are produced by training a random
forest upon one week's worth of data and testing it on a different
week's worth of data. In total, there are $8\times7=56$ entries.
The random forests had 30 trees like in the original analysis but
only had 3 separate 100,000-sample draws due to the smaller sample
set (one week, about 350,000 samples). All data for the selected training
session is used, and all testing data matching a participant in the
training set is tested with. Note that multiclass AUC is reported
both for pairs where training week happens before testing week, as
would be expected for an attacker, but also in pairs where testing
week happens before training week, which is relevant to motion re-identification
well after data collection.

The results in Figure \ref{fig:timing_delay} show a pattern that
multiclass AUC is higher when training and testing sessions have less
delay (i.e, at the near-diagonals) than when there is more delay (near
bottom left and top right). This effect varies somewhat across weeks,
but is still easily visible. A mixed-effect model predicting multiclass
AUC based on delay with random intercepts from training and testing
weeks is highly significant ($t(24.53)=-5.59,p=8.6\times10^{-6}$).

\subsection*{Identification features} \label{subsec:identification_features}

The features in this model were developed relative to \cite{Miller2020}
but have significant changes. In order to shed light on the relative
value of these change, we provide a series of intermediate models
and the relative changes in accuracy for each model, allowing the
reader to differentiate the contributions of each change to the feature
set. These are given in table \ref{tab:model}.

\begin{table}
\caption{Model Comparisons}
\label{tab:model} %
\begin{tabular}{llrrr}
\toprule 
Model  & Features  & Accuracy (C=232)  & Multiclass AUC  & C=30 Accuracy \tabularnewline
\midrule 
M1  & Miller et al. \cite{Miller2020}  & 7.76\%  & 75.18\%  & 21.20\% \tabularnewline
M2  & M1 without horizontal positions  & 25.29\%  & 81.91\%  & 43.10\% \tabularnewline
M3  & M2 without yaw  & 27.87\%  & 82.73\%  & 47.51\% \tabularnewline
M4  & M3 with windows of 3s, 10s, 30s  & 31.61\%  & 82.71\%  & 49.22\% \tabularnewline
M5  & M4 with body-space displacements  & 32.76\%  & 84.43\%  & 51.54\% \tabularnewline
M6  & M5 with speed values  & 32.47\%  & 85.59\%  & 52.59\% \tabularnewline
\bottomrule
\end{tabular}
\end{table}

Model M1 used the same feature set as used in \cite{Miller2020}.
Note that this is the lowest accuracy of all six models here. Models
M2 and M3 use a subset of features available to M1, specifically,
dropping the two horizontal positions and then dropping horizontal
orientation. These changes improve each evaluation metric substantially.
Note that the samples within the same session are not independent,
and in fact the samples highly correlated upon these features in particular.
Domain knowledge in leading these discussions and activities indicates
that the horizontal locations of a participant from session to session
varies almost arbitrarily. Disallowing these values allows other features
to be considered more strongly.

Model M4 quadruples the feature space by adding three new window sizes.
As the original work \cite{Miller2020} used only thirty-second clips,
while our work has on average about thirty minutes, we can increase
our window size without suffering the same loss in sample size. The
difference between M3 and M4 is puzzling, as accuracy goes up substantially
but multiclass AUC goes slightly down. We hypothesize this is due
to the larger feature space adding variance to predictions across
the board, but the larger time window providing an important distinguishing
factor when there are differences between just a handful of plausible
values.

M5 and M6 apply features from previous work. M5 adds features computed
from the body-coordinate displacement vectors $\mathbf{b}_{\alpha\beta}[f]$
described in the subsection on body-space coordinates, building on \cite{Miller2022a}.
This reintroduces the horizontal information lost in M2 and M3, but
this featurization is invariant to horizontal translations and rotations
of the participant. M6 adds velocity features for each stream, inspired
by \cite{Moore2021} that shows velocity is still identifiable, though
less so than position.

In summary, the use of body-space coordinates increases identification
both by removing poorly parameterized data and by reintroducing a
better parameterization. Additionally, we corroborate the value of
other featurizations on the current dataset.

\section*{Discussion}

This work continues to survey the risks that VR poses to privacy.
The most important question in this space is how identifying various
data sources, situations, and activities are, what makes these identifying,
and what can be done about it. By understanding what influences the
accuracy of de-anonymization techniques, researchers can develop more
effective and more efficient ways to limit risk to end users.

First, we show, in accord with other work \cite{Miller2022,Moore2021},
that the largest difference in identifiability is whether the enrollment
(training) and input (testing) sessions are delayed or adjacent. This
work has demonstrated this fact on the largest dataset to date in
terms of total recorded time. This finding is echoed in the timing
analysis, in which one minute from the same session is more identifying
than seven sessions of thirty minutes each. Overall, we infer that
there are several identifying variables at play, and some may work
on short time scales and some work on long time scales. Future work
ought not to look at one time scale but many.

Second, in response to previous work with varying identification sizes,
we select and justify the Multiclass AUC evaluation metric to evaluate
identifiability across sample sizes. A review of previous work indicated
a trend that simple accuracy-based measures of identifiability show
lower accuracy with larger sets of users to identify from. Removing
this variation can let future work clarify other important trends
in accuracy, such as time, feature selection, or activity.

Third, we investigate identifiability as participants have discussions
in VR. This activity is a very common activity that occurs in social
VR even without prompting, and it is far more available to attackers
because of its ubiquity. This obviates the need to have targets perform
an action based on trust of the authentication method (e.g., \cite{Kupin2019,Li2016}
or co-present social engineering in the VR environment (e.g., \cite{Falk2021}).

Finally, to better understand the risk of motion data, we develop
body-space coordinates to allow for the usage of horizontal motion
data. Both removing the reference to global coordinates and introducing
the reference to body-space coordinates increase identifiability.
This work takes a step forward in the arms race between identification
and de-identification.

Because of this work, we recommend software developers responsible
for this motion data to take several steps. First, developers ought
to protect this data with standard practices for personally identifying
data. When this data needs to be shared with others, it can be helpful
to reduce the time span available, minimize variation in activities,
or modify data to produce plausible deniability \cite{Nair2022a}.
Furthermore, there are developments in law that need to be made to
clarify the legal status of this data based on its risks to privacy
\cite{heller_watching_nodate}.

Some limitations of this work include that while participants knew
their motion data was collected, they were not aware what features
of their data would be most identifying so that they could change
their behavior to avoid being tracked, e.g. vary their height week-to-week
to fool the model. All participants used the same headset for the
entire duration of the study, which according to previous work \cite{MillerR2020,Miller2021}
can make identification easier. On the other hand, almost all participants
used a Meta Oculus Quest 2 headset, so idiosyncrasies across headsets
(e.g., tracking errors that appear differently in the Meta Oculus
Quest 2 headset vs. HTC Vive headset) could not be leveraged as partially
identifying. Finally, the activity performed (discussion) and the context (a college course) was selected
due to the authors' personal experiences with activities in social
VR, but a more formal study of social VR may indicate other activities
that commonly take place that an attacker may leverage.

The sampling process used to minimize discontinuities (see the subsection on duration) does introduce a bias in the sampling such
that the middle of the recorded time is more likely to be part of
the testing set than the training set.

Regarding attack models, some avenues for future work include demonstrating
effective attacks beyond biometrics. For example, depending on what
is transmitted, almost all of a target's visual and auditory experience
can be recorded or inferred by a VR lurker. This includes inferences
about the target's attention to objects, content, or people in the
virtual world coming from both conscious and unconscious mechanisms.
These features will likely involve models different from position-based
random forest, such as neural networks \cite{Miller2021}. On the
defensive side, there is still important work to be done on methods
of obfuscating motion data to minimize identifiability. Additionally,
it may be plausible, given a user's preferences, to disconnect real-world
biometrics like height and arm length from a user's virtual avatar
entirely, or use transformed social interaction \cite{Bailenson2004}
so that gestures that might otherwise be identifiable can come from
another person but still be communicative.


Ultimately we aim to see the users, designers, and developers of social
VR to understand and account for the risks of this medium so that
people can safely and consciously use this new and powerful medium.
\bibliography{sample}

\begin{thebibliography}{10}
\urlstyle{rm}
\expandafter\ifx\csname url\endcsname\relax
  \def\url#1{\texttt{#1}}\fi
\expandafter\ifx\csname urlprefix\endcsname\relax\def\urlprefix{URL }\fi
\expandafter\ifx\csname doiprefix\endcsname\relax\def\doiprefix{DOI: }\fi
\providecommand{\bibinfo}[2]{#2}
\providecommand{\eprint}[2][]{\url{#2}}

\bibitem{Yaremych2019}
\bibinfo{author}{Yaremych, H.~E.} \& \bibinfo{author}{Persky, S.}
\newblock \bibinfo{journal}{\bibinfo{title}{{Tracing physical behavior in
  virtual reality: A narrative review of applications to social psychology}}}.
\newblock {\emph{\JournalTitle{Journal of Experimental Social Psychology}}}
  \textbf{\bibinfo{volume}{85}}, \bibinfo{pages}{103845},
  \doiprefix\url{10.1016/j.jesp.2019.103845} (\bibinfo{year}{2019}).

\bibitem{Sweeney2000}
\bibinfo{author}{Sweeney, L.}
\newblock \bibinfo{title}{{Simple demographics often identify people uniquely}}
  (\bibinfo{year}{2000}).

\bibitem{Eckersley2010}
\bibinfo{author}{Eckersley, P.}
\newblock \bibinfo{journal}{\bibinfo{title}{{How unique is your web browser?}}}
\newblock {\emph{\JournalTitle{Lecture Notes in Computer Science (including
  subseries Lecture Notes in Artificial Intelligence and Lecture Notes in
  Bioinformatics)}}} \textbf{\bibinfo{volume}{6205 LNCS}},
  \bibinfo{pages}{1--18}, \doiprefix\url{10.1007/978-3-642-14527-8_1}
  (\bibinfo{year}{2010}).

\bibitem{Pfeuffer2019}
\bibinfo{author}{Pfeuffer, K.} \emph{et~al.}
\newblock \bibinfo{title}{{Behavioural Biometrics in VR: Identifying People
  from Body Motion and Relations in Virtual Reality}}.
\newblock In \emph{\bibinfo{booktitle}{Proceedings of the 2019 CHI Conference
  on Human Factors in Computing Systems}}, CHI '19,
  \bibinfo{pages}{110:1----110:12}, \doiprefix\url{10.1145/3290605.3300340}
  (\bibinfo{publisher}{ACM}, \bibinfo{address}{New York, NY, USA},
  \bibinfo{year}{2019}).

\bibitem{Moore2021}
\bibinfo{author}{Moore, A.~G.}, \bibinfo{author}{McMahan, R.~P.},
  \bibinfo{author}{Dong, H.} \& \bibinfo{author}{Ruozzi, N.}
\newblock \bibinfo{journal}{\bibinfo{title}{{Personal identifiability and
  obfuscation of user tracking data from VR training sessions}}}.
\newblock {\emph{\JournalTitle{Proceedings - 2021 IEEE International Symposium
  on Mixed and Augmented Reality, ISMAR 2021}}} \bibinfo{pages}{221--228},
  \doiprefix\url{10.1109/ISMAR52148.2021.00037} (\bibinfo{year}{2021}).

\bibitem{Miller2020}
\bibinfo{author}{Miller, M.~R.}, \bibinfo{author}{Herrera, F.},
  \bibinfo{author}{Jun, H.}, \bibinfo{author}{Landay, J.~A.} \&
  \bibinfo{author}{Bailenson, J.~N.}
\newblock \bibinfo{journal}{\bibinfo{title}{{Personal identifiability of user
  tracking data during observation of 360-degree VR video}}}.
\newblock {\emph{\JournalTitle{Scientific Reports}}}
  \textbf{\bibinfo{volume}{10}}, \bibinfo{pages}{17404--17413},
  \doiprefix\url{10.1038/s41598-020-74486-y} (\bibinfo{year}{2020}).

\bibitem{Nair2022}
\bibinfo{author}{Nair, V.}, \bibinfo{author}{Garrido, G.~M.} \&
  \bibinfo{author}{Song, D.}
\newblock \bibinfo{journal}{\bibinfo{title}{{Exploring the Unprecedented
  Privacy Risks of the Metaverse}}}.
\newblock {\emph{\JournalTitle{ArXiv}}}  (\bibinfo{year}{2022}).
\newblock \eprint{2207.13176}.

\bibitem{Miller2022}
\bibinfo{author}{Miller, R.}, \bibinfo{author}{Banerjee, N.~K.} \&
  \bibinfo{author}{Banerjee, S.}
\newblock \bibinfo{journal}{\bibinfo{title}{{Temporal Effects in Motion
  Behavior for Virtual Reality (VR) Biometrics}}}.
\newblock {\emph{\JournalTitle{Proceedings - 2022 IEEE Conference on Virtual
  Reality and 3D User Interfaces, VR 2022}}} \bibinfo{pages}{563--572},
  \doiprefix\url{10.1109/VR51125.2022.00076} (\bibinfo{year}{2022}).

\bibitem{Hand2001}
\bibinfo{author}{Hand, D.~J.} \& \bibinfo{author}{Till, R.~J.}
\newblock \bibinfo{journal}{\bibinfo{title}{{A Simple Generalisation of the
  Area Under the ROC Curve for Multiple Class Classification Problems}}}.
\newblock {\emph{\JournalTitle{Machine Learning}}}
  \textbf{\bibinfo{volume}{45}}, \bibinfo{pages}{171--186},
  \doiprefix\url{10.1023/A:1010920819831} (\bibinfo{year}{2001}).

\bibitem{Slater2020}
\bibinfo{author}{Slater, M.} \emph{et~al.}
\newblock \bibinfo{journal}{\bibinfo{title}{{The Ethics of Realism in Virtual
  and Augmented Reality}}}.
\newblock {\emph{\JournalTitle{Frontiers in Virtual Reality}}}
  \textbf{\bibinfo{volume}{1}}, \bibinfo{pages}{1--13},
  \doiprefix\url{10.3389/frvir.2020.00001} (\bibinfo{year}{2020}).

\bibitem{Bailenson2018}
\bibinfo{author}{Bailenson, J.}
\newblock \bibinfo{journal}{\bibinfo{title}{{Protecting Nonverbal Data Tracked
  in Virtual Reality}}}.
\newblock {\emph{\JournalTitle{JAMA Pediatrics}}}
  \textbf{\bibinfo{volume}{172}}, \bibinfo{pages}{905--906}
  (\bibinfo{year}{2018}).

\bibitem{Hosfelt2019}
\bibinfo{author}{Hosfelt, D.}
\newblock \bibinfo{title}{{Making ethical decisions for the immersive web}}
  (\bibinfo{year}{2019}).
\newblock \eprint{1905.06995}.

\bibitem{Roesner2014}
\bibinfo{author}{Roesner, F.}, \bibinfo{author}{Kohno, T.~O.} \&
  \bibinfo{author}{Molnar, D.}
\newblock \bibinfo{journal}{\bibinfo{title}{{Security and privacy for augmented
  reality systems}}}.
\newblock {\emph{\JournalTitle{Communications of the ACM}}}
  \textbf{\bibinfo{volume}{57}}, \bibinfo{pages}{88--96},
  \doiprefix\url{10.1145/2580723.2580730} (\bibinfo{year}{2014}).

\bibitem{Kupin2019}
\bibinfo{author}{Kupin, A.}, \bibinfo{author}{Moeller, B.},
  \bibinfo{author}{Jiang, Y.}, \bibinfo{author}{Banerjee, N.~K.} \&
  \bibinfo{author}{Banerjee, S.}
\newblock \bibinfo{title}{{Task-Driven Biometric Authentication of Users in
  Virtual Reality (VR) Environments}}.
\newblock In \emph{\bibinfo{booktitle}{Internation Conference on Multimedia
  Modeling2}}, \bibinfo{pages}{55--67} (\bibinfo{year}{2019}).

\bibitem{Shen2018}
\bibinfo{author}{Shen, Y.} \emph{et~al.}
\newblock \bibinfo{journal}{\bibinfo{title}{{GaitLock: Protect Virtual and
  Augmented Reality Headsets Using Gait}}}.
\newblock {\emph{\JournalTitle{IEEE Transactions on Dependable and Secure
  Computing}}} \textbf{\bibinfo{volume}{5971}}, \bibinfo{pages}{1--14},
  \doiprefix\url{10.1109/TDSC.2018.2800048} (\bibinfo{year}{2018}).

\bibitem{Li2016}
\bibinfo{author}{Li, S.} \emph{et~al.}
\newblock \bibinfo{journal}{\bibinfo{title}{{Whose move is it anyway?
  Authenticating smart wearable devices using unique head movement patterns}}}.
\newblock {\emph{\JournalTitle{2016 IEEE International Conference on Pervasive
  Computing and Communications, PerCom 2016}}} \bibinfo{pages}{1--9},
  \doiprefix\url{10.1109/PERCOM.2016.7456514} (\bibinfo{year}{2016}).

\bibitem{Olade2020}
\bibinfo{author}{Olade, I.}, \bibinfo{author}{Fleming, C.} \&
  \bibinfo{author}{Liang, H.~N.}
\newblock \bibinfo{journal}{\bibinfo{title}{{Biomove: Biometric user
  identification from human kinesiological movements for virtual reality
  systems}}}.
\newblock {\emph{\JournalTitle{Sensors (Switzerland)}}}
  \textbf{\bibinfo{volume}{20}}, \bibinfo{pages}{1--19},
  \doiprefix\url{10.3390/s20102944} (\bibinfo{year}{2020}).

\bibitem{Wang2021}
\bibinfo{author}{Wang, X.} \& \bibinfo{author}{Zhang, Y.}
\newblock \bibinfo{journal}{\bibinfo{title}{{Nod to Auth: Fluent AR/VR
  Authentication with User Head-Neck Modeling}}}.
\newblock {\emph{\JournalTitle{Conference on Human Factors in Computing Systems
  - Proceedings}}} \doiprefix\url{10.1145/3411763.3451769}
  (\bibinfo{year}{2021}).

\bibitem{Nair2022a}
\bibinfo{author}{Nair, V.}, \bibinfo{author}{Garrido, G.~M.} \&
  \bibinfo{author}{Song, D.}
\newblock \bibinfo{journal}{\bibinfo{title}{{Going Incognito in the
  Metaverse}}}.
\newblock {\emph{\JournalTitle{ArXiv}}}  (\bibinfo{year}{2022}).
\newblock \eprint{2208.05604}.

\bibitem{Falk2021}
\bibinfo{author}{Falk, B.}, \bibinfo{author}{Meng, Y.}, \bibinfo{author}{Zhan,
  Y.} \& \bibinfo{author}{Zhu, H.}
\newblock \bibinfo{journal}{\bibinfo{title}{{POSTER: ReAvatar: Virtual Reality
  De-anonymization Attack through Correlating Movement Signatures}}}.
\newblock {\emph{\JournalTitle{Proceedings of the ACM Conference on Computer
  and Communications Security}}} \bibinfo{pages}{2405--2407},
  \doiprefix\url{10.1145/3460120.3485345} (\bibinfo{year}{2021}).

\bibitem{sabra_exploiting_2023}
\bibinfo{author}{Sabra, M.}, \bibinfo{author}{Sureshkanth, N.~V.},
  \bibinfo{author}{Sharma, A.}, \bibinfo{author}{Maiti, A.} \&
  \bibinfo{author}{Jadliwala, M.}
\newblock \bibinfo{title}{Exploiting out-of-band motion sensor data to
  de-anonymize virtual reality users}.
\newblock \eprint{2301.09041 [cs]}.

\bibitem{Ajit2019}
\bibinfo{author}{Ajit, A.}, \bibinfo{author}{Banerjee, N.~K.} \&
  \bibinfo{author}{Banerjee, S.}
\newblock \bibinfo{journal}{\bibinfo{title}{{Combining pairwise feature matches
  from device trajectories for biometric authentication in virtual reality
  environments}}}.
\newblock {\emph{\JournalTitle{Proceedings - 2019 IEEE International Conference
  on Artificial Intelligence and Virtual Reality, AIVR 2019}}}
  \bibinfo{pages}{9--16}, \doiprefix\url{10.1109/AIVR46125.2019.00012}
  (\bibinfo{year}{2019}).

\bibitem{Liebers2021}
\bibinfo{author}{Liebers, J.} \emph{et~al.}
\newblock \bibinfo{journal}{\bibinfo{title}{Understanding user identification
  in virtual reality through behavioral biometrics and the efect of body
  normalization}}.
\newblock {\emph{\JournalTitle{Conference on Human Factors in Computing Systems
  - Proceedings}}} \doiprefix\url{10.1145/3411764.3445528}
  (\bibinfo{year}{2021}).
\newblock \bibinfo{note}{{ISBN}: 9781450380966}.

\bibitem{Han2023}
\bibinfo{author}{Han, E.}, \bibinfo{author}{Miller, M.~R.},
  \bibinfo{author}{Ram, N.}, \bibinfo{author}{Nowak, K.~L.} \&
  \bibinfo{author}{Bailenson, J.~N.}
\newblock \bibinfo{title}{Understanding group behavior in virtual reality: A
  large-scale, longitudinal study in the metaverse}.

\bibitem{Miller2022a}
\bibinfo{author}{Miller, R.}, \bibinfo{author}{Banerjee, N.~K.} \&
  \bibinfo{author}{Banerjee, S.}
\newblock \bibinfo{journal}{\bibinfo{title}{{Combining Real-World Constraints
  on User Behavior with Deep Neural Networks for Virtual Reality (VR)
  Biometrics}}}.
\newblock {\emph{\JournalTitle{Proceedings - 2022 IEEE Conference on Virtual
  Reality and 3D User Interfaces, VR 2022}}} \bibinfo{pages}{409--418},
  \doiprefix\url{10.1109/VR51125.2022.00060} (\bibinfo{year}{2022}).

\bibitem{RLang2021}
\bibinfo{author}{{R Core Team}}.
\newblock \emph{\bibinfo{title}{R: A Language and Environment for Statistical
  Computing}}.
\newblock \bibinfo{organization}{R Foundation for Statistical Computing},
  \bibinfo{address}{Vienna, Austria} (\bibinfo{year}{2021}).

\bibitem{rangerPkg2017}
\bibinfo{author}{Wright, M.~N.} \& \bibinfo{author}{Ziegler, A.}
\newblock \bibinfo{journal}{\bibinfo{title}{{ranger}: A fast implementation of
  random forests for high dimensional data in {C++} and {R}}}.
\newblock {\emph{\JournalTitle{Journal of Statistical Software}}}
  \textbf{\bibinfo{volume}{77}}, \bibinfo{pages}{1--17},
  \doiprefix\url{10.18637/jss.v077.i01} (\bibinfo{year}{2017}).

\bibitem{niculescu-mizil_predicting_2005}
\bibinfo{author}{Niculescu-Mizil, A.} \& \bibinfo{author}{Caruana, R.}
\newblock \bibinfo{title}{Predicting good probabilities with supervised
  learning}.
\newblock In \emph{\bibinfo{booktitle}{Proceedings of the 22nd international
  conference on Machine learning - {ICML} '05}}, \bibinfo{pages}{625--632},
  \doiprefix\url{10.1145/1102351.1102430} (\bibinfo{publisher}{{ACM} Press},
  \bibinfo{year}{2005}).

\bibitem{heller_watching_nodate}
\bibinfo{author}{Heller, B.}
\newblock \bibinfo{journal}{\bibinfo{title}{Watching androids dream of electric
  sheep: Immersive technology, biometric psychography, and the law}}.
\newblock {\emph{\JournalTitle{Vanderbilt Journal of Entertainment and
  Technology Law}}} \textbf{\bibinfo{volume}{23}} (\bibinfo{year}{2020}).

\bibitem{MillerR2020}
\bibinfo{author}{Miller, R.}, \bibinfo{author}{Banerjee, N.~K.} \&
  \bibinfo{author}{Banerjee, S.}
\newblock \bibinfo{journal}{\bibinfo{title}{{Within-System and Cross-System
  Behavior-Based Biometric Authentication in Virtual Reality}}}.
\newblock {\emph{\JournalTitle{Proceedings - 2020 IEEE Conference on Virtual
  Reality and 3D User Interfaces, VRW 2020}}} \bibinfo{pages}{311--316},
  \doiprefix\url{10.1109/VRW50115.2020.00070} (\bibinfo{year}{2020}).

\bibitem{Miller2021}
\bibinfo{author}{Miller, R.}, \bibinfo{author}{Banerjee, N.~K.} \&
  \bibinfo{author}{Banerjee, S.}
\newblock \bibinfo{journal}{\bibinfo{title}{{Using siamese neural networks to
  perform cross-system behavioral authentication in virtual reality}}}.
\newblock {\emph{\JournalTitle{Proceedings - 2021 IEEE Conference on Virtual
  Reality and 3D User Interfaces, VR 2021}}} \bibinfo{pages}{140--149},
  \doiprefix\url{10.1109/VR50410.2021.00035} (\bibinfo{year}{2021}).

\bibitem{Bailenson2004}
\bibinfo{author}{Bailenson, J.~N.}, \bibinfo{author}{Beall, A.~C.},
  \bibinfo{author}{Loomis, J.}, \bibinfo{author}{Blascovich, J.} \&
  \bibinfo{author}{Turk, M.}
\newblock \bibinfo{journal}{\bibinfo{title}{{Transformed social interaction:
  Decoupling representation from behavior and form in collaborative virtual
  environments}}}.
\newblock {\emph{\JournalTitle{Presence: Teleoperators and Virtual
  Environments}}} \textbf{\bibinfo{volume}{13}}, \bibinfo{pages}{428--441},
  \doiprefix\url{10.1162/1054746041944803} (\bibinfo{year}{2004}).

\end{thebibliography}

\section*{Acknowledgements}

The authors wish to thank Daniel Scott Akselrad, Tobin Asher, Brian Beams, Ryan Moore, Patricia Jeanne Yablonski, and Mark York for their support with the course. This research was supported by the National Science Foundation grant (Award 1800922).

\section*{Author contributions statement}

M.R.M. and J.N.B. conceived of the experiments. J.N.B. secured funding. M.R.M., E.H., and C.D. collected data. M.R.M., E.J., and R.C., developed software to analyze the data. M.R.M. drafted the manuscript. E.H., C.D., J.N.B, and E.J contributed to the manuscript. All authors reviewed the manuscript.

\section*{Data availability}
The datasets analysed during the current study are available from the corresponding author on reasonable request.

\section*{Additional information}

The author(s) declare no competing interests.

\section*{Supplemental Material}

\subsection*{Multiclass AUC}

To our knowledge, no work in the space of user identification with
VR data has used multiclass AUC. Because it addresses the effect of
classification size on accuracy, we give a short description and justification
of its use in enabling future comparisons across studies with varying
numbers of classes.

Identification-focused works \cite{Pfeuffer2019,Miller2020,Moore2021,Miller2022}
almost exclusively use accuracy for the model's evaluation metric.
The benefits of accuracy as a metric include its ease of interpretation
and its directness to the question at hand - a less accuracy model
is obviously less identifiable, and vice versa. However, accuracy
does vary significantly as the number of classes varies, even for
the same data distributions and identification processes, as evidenced
by multiple works \cite{Pfeuffer2019,Miller2020,Wang2021}. Intuitively,
this is true - it is easier to guess who is walking up the stairs
in an apartment with two other people than a house of ten. This effect
of the number of classes on accuracy can make synthesis of findings
across works difficult, as the classification can vary as much as
two orders of magnitude (e.g., 5 in \cite{Wang2021} to 511 in \cite{Miller2020}).

Our criteria for an evaluation metric that addresses this issue is
that it produces the same value regardless if it is computed upon
the full set of classes, or computed as the average of randomly chosen
subsets of classes of any size. More formally, let $\mathcal{C}$
represent a classification problem whose elements $C\in\mathcal{C}$
are sets containing individual members of the class $C$. We define
an ideal evaluation metric $\mathcal{M}$ such that the evaluation
$\mathcal{M}(f,\mathcal{C})$ computed from the prediction function
$f$ and the classification problem $\mathcal{C}$ is equal to the
expected value of the evaluation $\mathcal{M}(f,\mathcal{C}')$ for
a randomly chosen combination of classes $\mathcal{C}'$ of a given
size $N$, uniformly randomly selected from the classes in $\mathcal{C}$.
Numerically, this is:

\[
\mathcal{M}(f,\mathcal{C})=\binom{|\mathcal{C}|}{N}^{-1}\sum_{\mathcal{C}'\subseteq\mathcal{C},|\mathcal{C}'|=N}\mathcal{M}(f,\mathcal{C}')
\]

To solve this problem and enable comparisons across analyses with
varying numbers of classes, we choose our primary evaluation metric
to be \textit{multiclass AUC}, defined by Hand and Till \cite{Hand2001}.

Multiclass AUC can be described as the average of the pairwise separability
between classes. In the original work, Hand and Till extend area-under-the-curve
(AUC), the well-known measure of separability, to the multiclass case.
AUC can be expressed as the probability that a randomly selected member
$a$ of class $A$ will be larger than a randomly selected member
$b$ of class $B$ according to the value of the binary prediction
function $f_{binary}$ meant to separate the two. This can be easily
computed in closed form as 
\[
AUC=\frac{1}{|A||B|}\sum_{a\in A,\ b\in B}\mathbf{1}[f_{binary}(a)>f_{binary}(b)]
\]
where $\mathbf{1}$ is the indicator function. Multiclass AUC extends
this definition provided a multiclass prediction function $f$ that
specifies values $f(m,C)$ for each combination of member $m$ and
class $C$. From this, Hand and Till \cite{Hand2001} define the multiclass
AUC for a given prediction function $f$ and set of classes $\mathcal{C}$
to be the average of separabilities of one class from another for
all pairs of classes in the model:

\[
\frac{1}{|\mathcal{C}|(|\mathcal{C}|-1)}\sum_{A,B\in\mathcal{C},\ A\neq B}\frac{1}{|A||B|}\sum_{a\in A,\ b\in B}f(a,A)>f(b,A)
\]

As a sketch for the proof that metric fills the criteria above, consider
that this metric produces a separability value for each ordered pair
of classes independent of the other classes present. Due to the symmetry
of classes in being selected within the final set, each class and
class pair is weighted similarly in the averaging process. By linearity,
the average of the final values for a given class size can be understood
as the average of all pairwise values, which produces the same value
as evaluating the full set of classes.

Hand and Till note that this metric weights the separability of each
pair of classes equally regardless of the number of samples in the
classes, which may not be appropriate if priors are to be taken into
account. Additionally, this is not an estimate of the accuracy attained
by the same training process upon a smaller data set constructed in
the same class-reduction process, but is instead an estimate based
upon the model after training.

\subsection*{Accuracy limited to an $N$-class testing set}

While multiclass AUC is a good multiclass evaluation metric for future
work, there are no works in this space that currently use it. In order
to allow comparisons to be drawn from this work to previous work,
we define accuracy limited to N-classes. This metric may be narrated
as a prediction task in which there is a model and a set of N potential
classifications, a subset of all the classifications the model could
make. First, the model proposes its classification, and if the classification
is outside this subset, the model is asked to provide its next best
classification. This process only ends when the model gives a predicted
classification within the set of potential classifications. 

To derive this formula, consider the probability $P[\text{arg max}_{C\in\mathcal{C}'}f(a,C)=A]$
for a randomly selected subclassification $\mathcal{C}'\subset\mathcal{C}$,
$|\mathcal{C}'|=N<|\mathcal{C}|$, that sample $a$ known to be from
class $A$ is predicted correctly. Explained simply, for a sample
to be correctly classified, the random selection of classes within
this set of $N$ potential classifications must avoid all classes
that would trip up the prediction for a given sample $a$ whose true
class is $A$. The number of these 'error classes' is $N_{error}=\sum_{C\in\mathcal{C}}\mathbf{1}[f(a,C)>f(a,A)]$.
The general expression for a sample $a$ to be correctly classified
in an $N$-class testing set is a simple combinatorics expression:

\[
P[\arg\max_{C\in\mathcal{C}'}f(a,C)=A]=\frac{\binom{|\mathcal{C}|-N_{error}}{N}}{\binom{|\mathcal{C}|}{N}}
\]

Then, by linearity of expectation, the accuracy for the whole model
across all selections of $\mathcal{C}'\subset\mathcal{C}$ is equal
to the mean of each sample's accuracy, and so the end result is simply
the mean of the expression above across all sessions.

\end{document}